\documentclass[12pt,a4paper]{article}
\usepackage{amsmath}
\usepackage{latexsym}
\usepackage{amssymb}
\usepackage[dvips]{graphicx,color}
\makeatletter
\def\rddots{\mathinner{\mkern1mu\raise\p@%
    \vbox{\kern7\p@\hbox{.}}\mkern2mu%
    \raise4\p@\hbox{.}\mkern2mu\raise7\p@\hbox{.}\mkern1mu}}
\makeatother
\begin{document}

\title{\sl Beyond the Gaussian II : \\
A Mathematical Experiment}
\author{
  Kazuyuki FUJII
  \thanks{E-mail address : fujii@yokohama-cu.ac.jp }\quad and\ 
  Hiroshi OIKE
  \thanks{E-mail address : oike@tea.ocn.ne.jp }\\
  ${}^{*}$Department of Mathematical Sciences\\
  Yokohama City University\\
  Yokohama, 236--0027\\
  Japan\\
  ${}^{\dagger}$Takado\ 85--5,\ Yamagata, 990--2464\\
  Japan\\
  }
\date{}
\maketitle
\begin{abstract}
  This is a sequel to the paper [K. Fujii : SIGMA {\bf 7} (2011), 022, 12 pages]. 
  In this paper we treat a non-Gaussian integral based on a quartic polynomial 
  and make a mathematical experiment by use of MATHEMATICA 
  whether the integral is written in terms of its discriminant or not. 
\end{abstract}
%

%
%
%
%
\section{Introduction}
The {\bf Gaussian} is an abbreviation of all subjects related to 
the Gauss function $e^{-(px^{2}+qx+r)}$ like the Gaussian 
beam, Gaussian process, Gaussian noise, etc. 
It plays a fundamental role in Mathematics, Statistics, 
Physics and related disciplines. 

It is generally conceived that any attempts to generalize the Gaussian results
would meet formidable difficulties.
Hoping to overcome this high wall of difficulties of going beyond the Gaussian
in the near future, a first step was introduced in \cite{MS}, \cite{KF}. 

This is simply one modest step to go beyond the Gaussian but it already 
reveals many obstacles related with the big challenge of going further 
beyond the Gaussian.

\section{Cubic Case}
In this section we treat the cubic case. 
In the paper \cite{MS}\footnote{it is not easy for non--experts to 
understand this paper correctly} the following ``formula" is reported
\begin{equation}
\label{eq:cubic-integral}
\int\int e^{-\left(ax^{3}+bx^{2}y+cxy^{2}+dy^{3}\right)}dxdy
=
\frac{1}{\sqrt[6]{-D}}
\end{equation}
where $D$ is the discriminant of the cubic equation
\begin{equation}
ax^{3}+bx^{2}+cx+d=0,
\end{equation}
and it is given by
\begin{equation}
\label{eq:cubic-discriminant}
D=b^{2}c^{2}+18abcd-4ac^{3}-4b^{3}d-27a^{2}d^{2}.
\end{equation}
The formula (\ref{eq:cubic-integral}) is of course non-Gaussian. 
However, if we consider it in the framework of the real category  
then (\ref{eq:cubic-integral}) is not correct because the left hand side 
diverges. In this paper we treat only the real category, and so $a, b, c, d, 
x, y$ are real numbers.

Formally, by performing the change of variable \ \ $x=t\rho,\ y=\rho$\ \ 
for (\ref{eq:cubic-integral}) we have
\begin{eqnarray*}
\mbox{LHS of (\ref{eq:cubic-integral})}
&=&
\int\int e^{-\rho^{3}\left(at^{3}+bt^{2}+ct+d\right)}|\rho|dtd\rho \\
&=&
\int
\left\{\int e^{-\left(at^{3}+bt^{2}+ct+d\right)\rho^{3}}|\rho|d\rho\right\}dt 
\\
&=&
\int |\sigma|e^{-\sigma^{3}}d\sigma
\int \frac{1}{|\sqrt[3]{(at^{3}+bt^{2}+ct+d)}|\sqrt[3]{(at^{3}+bt^{2}+ct+d)}}dt
\end{eqnarray*}
by the change of variable \ $\sigma=\sqrt[3]{at^{3}+bt^{2}+ct+d}\ \rho$. 

The divergence comes from
\[
\int |\sigma|e^{-\sigma^{3}}d\sigma,
\]
while the main part is 
\[
\int \frac{1}{|\sqrt[3]{(ax^{3}+bx^{2}+cx+d)}|\sqrt[3]{(ax^{3}+bx^{2}+cx+d)}}dx
\]
under the change $t\rightarrow x$. As a kind of renormalization 
the integral may be defined like
\[
\ddagger \int\int_{{\bf R}^{2}}e^{-(ax^{3}+bx^{2}y+cxy^{2}+dy^{3})}dxdy\ \ddagger 
=
\int_{{\bf R}}
\frac{1}{|\sqrt[3]{(ax^{3}+bx^{2}+cx+d)}|\sqrt[3]{(ax^{3}+bx^{2}+cx+d)}}dx.
\]

\noindent
However, the right hand side lacks proper symmetry. If we set
\begin{equation}
\label{eq:finite-integral}
F(a,b,c,d)=\int\int_{D_{R}}e^{-\left(ax^{3}+bx^{2}y+cxy^{2}+dy^{3}\right)}dxdy
\end{equation}
where $D_{R}=[-R,R]\times [-R,R]\ (R\gg 0)$, then it is easy to see
\[
F(-a,-b,-c,-d)=F(a,b,c,d).
\]
Namely, F is invariant under ${\bf Z}_{2}$--action.  This symmetry is 
important and must be kept even in the renormalization process. 
The right hand side in the ``definition" above is clearly not invariant. 
Therefore, by modifying it slightly we reach the renormalized integral

\vspace{3mm} \noindent
{\bf Definition}
\begin{equation}
\label{eq:renorm-definition}
\ddagger \int\int_{{\bf R}^{2}}e^{-(ax^{3}+bx^{2}y+cxy^{2}+dy^{3})}dxdy\ \ddagger 
=
\int_{{\bf R}} \frac{1}{\sqrt[3]{(ax^{3}+bx^{2}+cx+d)^{2}}}dx.
\end{equation}

\vspace{3mm}
The Gamma--function $\Gamma(p)$ is defined by
\begin{equation}
\label{eq:Gamma-function}
\Gamma(p)=\int_{0}^{\infty}e^{-x}x^{p-1}dx\quad (p>0)
\end{equation}
and the Beta--function $B(p,q)$ is
\begin{equation}
\label{eq:Beta-function}
B(p,q)=\int_{0}^{1}x^{p-1}(1-x)^{q-1}dx\quad (p,\ q>0).
\end{equation}
Then the result in \cite{KF} is

\vspace{5mm}
\begin{Large}
\noindent
{\bf Formula}\\
\end{Large}

\vspace{-5mm}
\noindent
(I)\ For $D < 0$ 
\begin{equation}
\label{eq:formula-I}
\int_{{\bf R}} \frac{1}{\sqrt[3]{(ax^{3}+bx^{2}+cx+d)^{2}}}dx
=
\frac{C_{-}}{\sqrt[6]{-D}}
\end{equation}
where
\[
C_{-}=\sqrt[3]{2}B(\frac{1}{2},\frac{1}{6}).
\]

\vspace{5mm}
\noindent
(II)\ For $D > 0$ 
\begin{equation}
\label{eq:formula-II}
\int_{{\bf R}} \frac{1}{\sqrt[3]{(ax^{3}+bx^{2}+cx+d)^{2}}}dx
=
\frac{C_{+}}{\sqrt[6]{D}}
\end{equation}
where
\[
C_{+}=3B(\frac{1}{3},\frac{1}{3}).
\]

\vspace{5mm}
\noindent
(III)\ $C_{-}$ and $C_{+}$ are related by $C_{+}=\sqrt{3}C_{-}$ 
through the identity
\begin{equation}
\label{eq:formula-III}
\sqrt{3}B(\frac{1}{3},\frac{1}{3})=\sqrt[3]{2}B(\frac{1}{2},\frac{1}{6}).
\end{equation}

See \cite{KF} in detail. 
Our result shows that the integral depends on the sign of $D$. 
This formula has been conjectured by Morozov and Shakirov \cite{MS} 
in a different context.

\vspace{3mm}
A comment is in order. \ If we treat the Gaussian case 
(: $e^{-(ax^{2}+bxy+cy^{2})}$) then the integral is reduced to
\begin{equation}
\label{eq:gaussian-formula}
\int_{{\bf R}}\frac{1}{ax^{2}+bx+c}dx
=\frac{2\pi}{\sqrt{-D}}
=\frac{2B(\frac{1}{2},\frac{1}{2})}{\sqrt{-D}}
\end{equation}
if $a>0$ and $D=b^{2}-4ac<0$. Because
\[
\pi=\frac{\sqrt{\pi}\sqrt{\pi}}{1}=
\frac{\Gamma(\frac{1}{2})\Gamma(\frac{1}{2})}{\Gamma(1)}=
B(\frac{1}{2},\frac{1}{2}).
\]

\vspace{5mm}
Let us check whether the renormalized integral  
(\ref {eq:renorm-definition}) is reasonable or not 
by making use of the results.

For the integral (\ref{eq:finite-integral}) it is easy to see
\begin{small}
\begin{eqnarray}
\hspace{-20mm}
\label{eq:identity-1}
\left(
\frac{\partial}{\partial a}\frac{\partial}{\partial d}-
\frac{\partial}{\partial b}\frac{\partial}{\partial c}
\right)F(a,b,c,d)
&=&
\int\int_{D_{R}}(x^{3}\cdot y^{3}-x^{2}y\cdot xy^{2})
e^{-\left(ax^{3}+bx^{2}y+cxy^{2}+dy^{3}\right)}dxdy
=0, \nonumber \\
\left(
\frac{\partial}{\partial b}\frac{\partial}{\partial b}-
\frac{\partial}{\partial a}\frac{\partial}{\partial c}
\right)F(a,b,c,d)
&=&
\int\int_{D_{R}}(x^{2}y\cdot x^{2}y-x^{3}\cdot xy^{2})
e^{-\left(ax^{3}+bx^{2}y+cxy^{2}+dy^{3}\right)}dxdy
=0,  \nonumber \\
&{}& \\
\left(
\frac{\partial}{\partial c}\frac{\partial}{\partial c}-
\frac{\partial}{\partial b}\frac{\partial}{\partial d}
\right)F(a,b,c,d)
&=&
\int\int_{D_{R}}(xy^{2}\cdot xy^{2}-x^{2}y\cdot y^{3})
e^{-\left(ax^{3}+bx^{2}y+cxy^{2}+dy^{3}\right)}dxdy
=0.  \nonumber
\end{eqnarray}
\end{small}

On the other hand, if we set
\[
{\cal F}(a,b,c,d)
=\int_{{\bf R}} \frac{1}{\sqrt[3]{(ax^{3}+bx^{2}+cx+d)^{2}}}dx
=\frac{C_{\pm}}{\sqrt[6]{\pm D}},
\]
then we can also verify the same relations
\begin{eqnarray}
\label{eq:identity-2}
&&\left(
\frac{\partial}{\partial a}\frac{\partial}{\partial d}-
\frac{\partial}{\partial b}\frac{\partial}{\partial c}
\right){\cal F}(a,b,c,d)=0, \nonumber \\
&&\left(
\frac{\partial}{\partial b}\frac{\partial}{\partial b}-
\frac{\partial}{\partial a}\frac{\partial}{\partial c}
\right){\cal F}(a,b,c,d)=0, \\
&&\left(
\frac{\partial}{\partial c}\frac{\partial}{\partial c}-
\frac{\partial}{\partial b}\frac{\partial}{\partial d}
\right){\cal F}(a,b,c,d)=0. \nonumber
\end{eqnarray}
by use of MATHEMATICA.

\vspace{3mm}
Therefore we can say that the definition (\ref{eq:renorm-definition}) 
is not so bad (maybe, good).

\section{Quartic Case : Mathematical Experiment}
In this section we treat the quartic case. The discriminant of 
the quartic equation
\begin{equation}
ax^{4}+bx^{3}+cx^{2}+dx+e=0
\end{equation}
is given by
\begin{eqnarray}
\label{eq:quartic-discriminant}
D
&=&
256a^{3}e^{3}-4b^{3}d^{3}-27a^{2}d^{4}-27b^{4}e^{2}-128a^{2}c^{2}e^{2}
+b^{2}c^{2}d^{2}+16ac^{4}e  \nonumber \\
&&
-4ac^{3}d^{2}-4b^{2}c^{3}e+144a^{2}cd^{2}e-6ab^{2}d^{2}e+144ab^{2}ce^{2}
-192a^{2}bde^{2}  \nonumber \\
&&
+18abcd^{3}+18b^{3}cde-80abc^{2}de.
\end{eqnarray}
See for example \cite{KF}.

Here we consider a non--Gaussian integral
\begin{equation}
\label{eq:quartic-integral}
{\cal F}\equiv {\cal F}(a,b,c,d,e)=
\int\int e^{-\left(ax^{4}+bx^{3}y+cx^{2}y^{2}+dxy^{3}+ey^{4}\right)}dxdy
\end{equation}
and study whether this integral can be written in terms of 
its discriminant (\ref{eq:quartic-discriminant}) or not like the 
formula in the preceding section. 
The conclusion is {\bf negative}, while we have very interesting 
mathematical ``phenomena" stated in the following.

By use of the same change of variable \ \ $x=t\rho,\ y=\rho$\ \ 
in the cubic case we have
\[
{\cal F}
=
\int\int e^{-\rho^{4}\left(at^{4}+bt^{3}+ct^{2}+dt+e\right)}|\rho|dtd\rho \\
=
\int
\left\{\int e^{-\left(at^{4}+bt^{3}+ct^{2}+dt+e\right)\rho^{4}}|\rho|d\rho\right\}dt 
\]
and the change of variable 
$\sigma=\sqrt[4]{at^{4}+bt^{3}+ct^{2}+dt+e}\rho$ ($at^{4}+bt^{3}+ct^{2}+dt+e>0$) 
gives
\begin{eqnarray}
\label{eq:reduced integral}
{\cal F}
&=&
\int |\sigma|e^{-\sigma^{4}}d\sigma
\int \frac{1}{\sqrt[4]{(at^{4}+bt^{3}+ct^{2}+dt+e)^{2}}}dt \nonumber \\
&=&
\frac{\sqrt{\pi}}{2}
\int_{{\bf R}}\frac{1}{\sqrt[4]{(ax^{4}+bx^{3}+cx^{2}+dx+e)^{2}}}dx
\end{eqnarray}
because
\[
\int_{-\infty}^{\infty}|\sigma|e^{-\sigma^{4}}d\sigma
=2\int_{0}^{\infty}\sigma e^{-\sigma^{4}}d\sigma
=\frac{1}{2}\Gamma\left(\frac{1}{2}\right)
=\frac{\sqrt{\pi}}{2}.
\]

\vspace{3mm}\noindent
{\bf Problem I}\ \ How can we calculate (\ref{eq:reduced integral}) ?

\vspace{3mm}
There is no method to calculate at the present time, so we make 
a mathematical experiment by use of MATHEMATICA. 
From the lesson of quadratic and cubic cases it may conjecture
\begin{equation}
{\cal F}=\frac{C}{\sqrt[12]{-D}}\quad (D<0,\ a>0)
\end{equation}
where $C$ is some constant and $12=4\times 3$.

\vspace{3mm}
For the integral of exponent (\ref{eq:quartic-integral}) we have 
a system of differential equations
\begin{eqnarray}
\label{eq:Identity-1}
\left(
\frac{\partial}{\partial {a}}\frac{\partial}{\partial {c}}-
\frac{\partial^{2}}{{\partial {b}}^{2}}
\right) {\cal F}&=&0, \nonumber \\
\left(
\frac{\partial}{\partial {a}}\frac{\partial}{\partial {d}}-
\frac{\partial}{\partial {b}}\frac{\partial}{\partial {c}}
\right) {\cal F}&=&0, \nonumber \\
\left(
\frac{\partial}{\partial {a}}\frac{\partial}{\partial {e}}-
\frac{\partial^{2}}{{\partial {c}}^{2}}
\right) {\cal F}&=&0, \nonumber \\
\left(
\frac{\partial}{\partial {b}}\frac{\partial}{\partial {d}}-
\frac{\partial^{2}}{{\partial {c}}^{2}}
\right) {\cal F}&=&0, \\
\left(
\frac{\partial}{\partial {b}}\frac{\partial}{\partial {e}}-
\frac{\partial}{\partial {c}}\frac{\partial}{\partial {d}}
\right) {\cal F}&=&0, \nonumber \\
\left(
\frac{\partial}{\partial {c}}\frac{\partial}{\partial {e}}-
\frac{\partial^{2}}{{\partial {d}}^{2}}
\right) {\cal F}&=&0. \nonumber 
\end{eqnarray}
The proof is straightforward. For example, 
\[
\left(
\frac{\partial}{\partial {a}}\frac{\partial}{\partial {c}}-
\frac{\partial^{2}}{{\partial {b}}^{2}}
\right) {\cal F}
=
\int\int \left\{x^{4}\cdot x^{2}y^{2}-(x^{3}y)^{2}\right\}
e^{-\left(ax^{4}+bx^{3}y+cx^{2}y^{2}+dxy^{3}+ey^{4}\right)}dxdy
=0.
\]

On the other hand, if we set
\begin{equation}
\widetilde{{\cal F}}=\frac{C}{\sqrt[12]{-D}}
\end{equation}
then we have
\begin{eqnarray}
\label{eq:Identity-2}
\left(
\frac{\partial}{\partial {a}}\frac{\partial}{\partial {c}}-
\frac{\partial^{2}}{{\partial {b}}^{2}}
\right)\widetilde{{\cal F}}&=&-C
\frac{(c^{2}-3bd+12ae)(3d^{2}-8ce)}{36(-D)^{13/12}}, 
\nonumber \\
\left(
\frac{\partial}{\partial {a}}\frac{\partial}{\partial {d}}-
\frac{\partial}{\partial {b}}\frac{\partial}{\partial {c}}
\right)\widetilde{{\cal F}}&=&C
\frac{(c^{2}-3bd+12ae)(cd-6be)}{18(-D)^{13/12}}, 
\nonumber \\
\left(
\frac{\partial}{\partial {a}}\frac{\partial}{\partial {e}}-
\frac{\partial^{2}}{{\partial {c}}^{2}}
\right)\widetilde{{\cal F}}&=&-C
\frac{(c^{2}-3bd+12ae)(c^{2}-2bd-4ae)}{9(-D)^{13/12}}, 
\nonumber \\
\left(
\frac{\partial}{\partial {b}}\frac{\partial}{\partial {d}}-
\frac{\partial^{2}}{{\partial {c}}^{2}}
\right)\widetilde{{\cal F}}&=&C
\frac{(c^{2}-3bd+12ae)(16ae-bd)}{36(-D)^{13/12}}, 
\\
\left(
\frac{\partial}{\partial {b}}\frac{\partial}{\partial {e}}-
\frac{\partial}{\partial {c}}\frac{\partial}{\partial {d}}
\right)\widetilde{{\cal F}}&=&-C
\frac{(c^{2}-3bd+12ae)(6ad-bc)}{18(-D)^{13/12}}, 
\nonumber \\
\left(
\frac{\partial}{\partial {c}}\frac{\partial}{\partial {e}}-
\frac{\partial^{2}}{{\partial {d}}^{2}}
\right)\widetilde{{\cal F}}&=&-C
\frac{(c^{2}-3bd+12ae)(3b^{2}-8ac)}{36(-D)^{13/12}} 
\nonumber 
\end{eqnarray}
by use of MATHEMATICA (verification by hand is very tough).

\vspace{3mm}
As a result ${\cal F} \ne \widetilde{{\cal F}}$. However, 
from (\ref{eq:Identity-2}) we obtain an interesting quantity
\begin{equation}
E\equiv c^{2}-3bd+12ae\ \Longleftarrow\  
ax^{4}+bx^{3}+cx^{2}+dx+e.
\end{equation}
It is not clear at the present time what $E$ is, so we present

\vspace{3mm}\noindent
{\bf Problem II}\ \ Make the property of $E$ clear.

\vspace{3mm}\noindent
We believe that $E$ will play an important role in the calculation.

\vspace{3mm}
In last, we note some interesting fact. If we set
\[
\widehat{{\cal F}}=\frac{C}{\sqrt[12]{D}}\quad (D>0)
\]
then the same relations (\ref{eq:Identity-2}) hold 
apart from the sign

\begin{eqnarray}
\left(
\frac{\partial}{\partial {a}}\frac{\partial}{\partial {c}}-
\frac{\partial^{2}}{{\partial {b}}^{2}}
\right)\widehat{{\cal F}}&=&C
\frac{(c^{2}-3bd+12ae)(3d^{2}-8ce)}{36D^{13/12}}, 
\nonumber \\
\left(
\frac{\partial}{\partial {a}}\frac{\partial}{\partial {d}}-
\frac{\partial}{\partial {b}}\frac{\partial}{\partial {c}}
\right)\widehat{{\cal F}}&=&-C
\frac{(c^{2}-3bd+12ae)(cd-6be)}{18D^{13/12}}, 
\nonumber \\
\left(
\frac{\partial}{\partial {a}}\frac{\partial}{\partial {e}}-
\frac{\partial^{2}}{{\partial {c}}^{2}}
\right)\widehat{{\cal F}}&=&C
\frac{(c^{2}-3bd+12ae)(c^{2}-2bd-4ae)}{9D^{13/12}}, 
\nonumber \\
\left(
\frac{\partial}{\partial {b}}\frac{\partial}{\partial {d}}-
\frac{\partial^{2}}{{\partial {c}}^{2}}
\right)\widehat{{\cal F}}&=&-C
\frac{(c^{2}-3bd+12ae)(16ae-bd)}{36D^{13/12}}, 
\nonumber \\
\left(
\frac{\partial}{\partial {b}}\frac{\partial}{\partial {e}}-
\frac{\partial}{\partial {c}}\frac{\partial}{\partial {d}}
\right)\widehat{{\cal F}}&=&C
\frac{(c^{2}-3bd+12ae)(6ad-bc)}{18D^{13/12}}, 
\nonumber \\
\left(
\frac{\partial}{\partial {c}}\frac{\partial}{\partial {e}}-
\frac{\partial^{2}}{{\partial {d}}^{2}}
\right)\widehat{{\cal F}}&=&C
\frac{(c^{2}-3bd+12ae)(3b^{2}-8ac)}{36D^{13/12}}.
\nonumber 
\end{eqnarray}

These ``phenomena" are interesting enough and 
worth studying in detail.

\section{Concluding Remarks}
In this note we treated a non-Gaussian integral based on a quartic 
polynomial and made a mathematical experiment by use of MATHEMATICA. 
Though our work is far from obtaining an explicit value of the integral, we 
found an interesting quantity $E$. In order to make it clear hard work will be 
needed. 

See \cite{MS2}, \cite{AS} as recent results on this topic and \cite{DM} 
as a general introduction to non--linear algebras. 
We expect young mathematicians or mathematical physicists to take part in 
this fascinating topic.

\vspace{10mm}
\noindent{\em Acknowledgment.}\\
The author wishes to thank Ryu Sasaki for helpful comments and 
suggestions.


%

\end{document}